%% file: main.tex
\documentclass[12pt,a4paper]{article}
\usepackage[utf8]{inputenc}

\usepackage[colorinlistoftodos]{todonotes}

\input{symbols_def.tex}

\begin{document}
\maketitle

\input{abstract.tex}

\input{intro.tex}

\input{mixing_indCPV.tex}

\input{directCPV.tex}
\input{Todd.tex}

\input{baryons.tex}
\input{prospects.tex}
\input{conclusion.tex}

\input{reference.tex}
\end{document}

%% file: symbols_def.tex
\def\paperauthors{Vishal Bhardwaj, Mirco Dorigo, Fu-Sheng Yu} 
\def\paperasciititle{CKM 2018: summary of WG7} 
\def\papertitle{\bf Summary of WG7 at CKM 2018: \\ ``Mixing and \boldmath{\CP} violation in the \boldmath{$D$} system: \boldmath{$x_D$}, \boldmath{$y_D$}, \boldmath{$|q/p|_D$}, \boldmath{$\phi_D$}, and direct \\ \boldmath{\CP} violation  in \boldmath{$D$} decays''} 
\def\paperkeywords{{High Energy Physics}, {Flavour Physics}, {Charm Physics}} 

\def\paperlicenceurl{https://creativecommons.org/licenses/by/4.0/}

\usepackage{authblk}
\title{\papertitle}
\author[a]{Vishal Bhardwaj}
\author[b]{Mirco Dorigo}
\author[c]{Fu-Sheng Yu}
\affil[a]{Indian Institute of Science Education and Research Mohali, India}
\affil[b]{CERN and INFN Sezione di Trieste}
\affil[c]{School of Nuclear Science and Technology, Lanzhou University, China}
\date{\today}                   
\usepackage[top=1in, bottom=1.25in, left=1in, right=1in]{geometry}
\usepackage{microtype}
\usepackage{lineno}
\usepackage{xspace}
\usepackage{booktabs}
\usepackage{comment}

\usepackage{caption} 

\usepackage{graphicx}  
\usepackage{color}
\usepackage{colortbl}
\DeclareGraphicsExtensions{.pdf,.PDF,png,.PNG}

\usepackage{amsmath} 
\usepackage{amssymb}
\usepackage{amsfonts}
\usepackage{upgreek} 

\usepackage{hyperxmp}
\usepackage[pdftex,
            pdfauthor={\paperauthors},
            pdftitle={\paperasciititle},
            pdfkeywords={\paperkeywords},
            pdflicenseurl={\paperlicenceurl}]{hyperref}
\usepackage[all]{hypcap} 

\def\Journal#1#2#3#4{{#1} {\bf #2}, #3 (#4)}

\def\PLB{{ Phys. Lett.}  B}
\def\PRL{ Phys. Rev. Lett.}
\def\PRD{{ Phys. Rev.} D}

\def\etal{{\textit{et al.}}}
\def\Belle{{Belle Collaboration}}
\def\BaBar{{BaBar Collaboration}}
\def\LHCb{{LHCb Collaboration}}
\def\BESIII{{BESIII Collaboration}}

\def\CP      {{\ensuremath{C\!P}}\xspace}
\def\T      {{\ensuremath{T}}\xspace}
\def\atodd   {{\ensuremath{a_{\CP}^{\rm\T-odd}}}\xspace}
\newcommand{\ket}[1]{\ensuremath{|#1\rangle}}

\def\muon   {{\ensuremath{\mu}}\xspace}
\def\mup    {{\ensuremath{\muon^+}}\xspace}
\def\mum    {{\ensuremath{\muon^-}}\xspace}

\def\pion   {{\ensuremath{\pi}}\xspace}
\def\piz    {{\ensuremath{\pion^0}}\xspace}
\def\pip    {{\ensuremath{\pion^+}}\xspace}
\def\pim    {{\ensuremath{\pion^-}}\xspace}
\def\pipm   {{\ensuremath{\pion^\pm}}\xspace}

\def\kaon   {{\ensuremath{K}}\xspace}
\def\Kbar    {{\kern 0.2em\overline{\kern -0.2em \kaon}{}}\xspace}

\def\KorKbar    {\kern 0.18em\optbar{\kern -0.18em K}{}\xspace}

\def\Kp      {{\ensuremath{\kaon^+}}\xspace}
\def\Km      {{\ensuremath{\kaon^-}}\xspace}

\def\Kmp     {{\ensuremath{\kaon^\mp}}\xspace}
\def\KS      {{\ensuremath{\kaon^0_{\mathrm{ \scriptscriptstyle S}}}}\xspace}
\def\KL      {{\ensuremath{\kaon^0_{\mathrm{ \scriptscriptstyle L}}}}\xspace}

\def\Dbar    {{\kern 0.2em\overline{\kern -0.2em D}{}}\xspace}
\def\D       {{\ensuremath{D}}\xspace}

\def\DorDbar    {\kern 0.18em\optbar{\kern -0.18em D}{}\xspace}
\def\Dz      {{\ensuremath{\D^0}}\xspace}
\def\Dzb     {{\ensuremath{\Dbar{}^0}}\xspace}
\def\Dp      {{\ensuremath{\D^+}}\xspace}

\def\Dstarp  {{\ensuremath{\D^{*+}}}\xspace}

\def\Xires       {{\ensuremath{\Xi}}\xspace}

\def\Lz          {{\ensuremath{\Lambda}}\xspace}
\def\Lbar        {{\ensuremath{\kern 0.1em\overline{\kern -0.1em\Lambda}}}\xspace}
\def\LorLbar    {\kern 0.18em\optbar{\kern -0.18em \PLambda}{}\xspace}

\def\Omegares    {{\ensuremath{\Omega}}\xspace}

\def\Lb      {{\ensuremath{\Lz^0_b}}\xspace}

\def\Lc      {{\ensuremath{\Lz^+_c}}\xspace}

\def\Xicz    {{\ensuremath{\Xires^0_c}}\xspace}
\def\Xicp    {{\ensuremath{\Xires^+_c}}\xspace}

\def\Omegac    {{\ensuremath{\Omegares^0_c}}\xspace}

\def\Omegab    {{\ensuremath{\Omegares^-_b}}\xspace}

\def\Xiccpp    {{\ensuremath{\Xires^{++}_{cc}}}\xspace}

\newcommand{\tev}{\ensuremath{\mathrm{\,Te\kern -0.1em V}}\xspace}
\newcommand{\gev}{\ensuremath{\mathrm{\,Ge\kern -0.1em V}}\xspace}
\newcommand{\mev}{\ensuremath{\mathrm{\,Me\kern -0.1em V}}\xspace}
\newcommand{\kev}{\ensuremath{\mathrm{\,ke\kern -0.1em V}}\xspace}
\newcommand{\ev}{\ensuremath{\mathrm{\,e\kern -0.1em V}}\xspace}
\newcommand{\gevc}{\ensuremath{{\mathrm{\,Ge\kern -0.1em V\!/}c}}\xspace}
\newcommand{\mevc}{\ensuremath{{\mathrm{\,Me\kern -0.1em V\!/}c}}\xspace}
\newcommand{\gevcc}{\ensuremath{{\mathrm{\,Ge\kern -0.1em V\!/}c^2}}\xspace}
\newcommand{\gevgevcccc}{\ensuremath{{\mathrm{\,Ge\kern -0.1em V^2\!/}c^4}}\xspace}
\newcommand{\mevcc}{\ensuremath{{\mathrm{\,Me\kern -0.1em V\!/}c^2}}\xspace}

\def\invpb {\ensuremath{\mbox{\,pb}^{-1}}\xspace}

\def\invfb   {\ensuremath{\mbox{\,fb}^{-1}}\xspace}

\def\invab   {\ensuremath{\mbox{\,ab}^{-1}}\xspace}

\usepackage{cite} 
\usepackage{mciteplus}

%% file: abstract.tex
\begin{abstract}
The {10$^{\rm th}$ \it International Workshop on the CKM Unitary Triangle} took place at the University of Heidelberg on September 17$^{\rm th}$--21$^{\rm st}$, 2018. In this write-up, we summarize the material discussed at the workshop by the Working Group 7, which focused on latest experimental results and theoretical developments in the study of mixing and \CP violation in the neutral $D$ system, and of \CP violation and decay properties of other charm mesons and baryons.
\end{abstract}

%% file: intro.tex
\section{Introduction}
The study of \Dz-\Dzb mixing offers an unique opportunity to probe flavour changing neutral currents between up-type quarks, which could be affected by physics beyond the standard model (SM) in very different ways compared to the down-type quarks featuring the neutral strange and bottom mesons systems. Violation of the \CP symmetry in mixing-induced processes is predicted to be very small in the SM, giving good prospects for the indirect observation of non-SM contributions. Effects of \CP violation in decays of charm mesons and baryons are difficult to predict; nonetheless, the current level of the most precise measurements, around $10^{-3}$, is getting closer to that needed for the observation of the tiny \CP asymmetries generally expected in the SM.  In these proceedings, we summarize the most recent efforts in this endeavour which were discussed by the Working Group 7 at the last {\it International Workshop on the CKM Unitary Triangle}, held at the University of Heidelberg on September 17$^{\rm th}$--21$^{\rm st}$, 2018. 

We divide the discussion into the following parts. Mixing and indirect \CP violation studies are reported in Sect.~\ref{sect:mixing_indCPV}; direct \CP violation is discussed in Sect.~\ref{sect:direct_CPV}. A brief report on \CP-violating triple product asymmetries is presented in Sect.~\ref{sect:TP_asym}, while Sect.~\ref{sect:baryons} reports results on charmed baryons, a topic which became very interesting recently. We present some experimental prospects in Sect.~\ref{sect:prospects}, and give a conclusion in Sect.~\ref{sect:conclusion}.   Charge-conjugation is implied in what follows, and when more than one uncertainty is assigned to a result, the first refers to the statistical uncertainty, the second to the systematic. These conventions are applied through the entire document, unless stated otherwise.

%% file: mixing_indCPV.tex
\section{Mixing and indirect \boldmath{\CP} violation}
\label{sect:mixing_indCPV}

The physical states of the time-dependent Hamiltonian describing the neutral $D$ system are superpositions of the flavor eigenstates, ${\ket{D_{1,2}} = p\ket{\Dz} \pm q \ket{\Dzb}}$, and this generates flavor mixing. The parameters $p$ and $q$ are complex numbers which satisfy the relation $|p|^2+|q|^2=1$. If charge-parity (\CP) symmetry holds, $q$ is equal to $p$ and two dimensionless parameters, $x \equiv  (m_1 - m_2)/\Gamma$ and $y \equiv (\Gamma_1 - \Gamma_2)/ 2\Gamma$, fully describe the oscillations. Here, $m_{1(2)}$ and $\Gamma_{1(2)}$ are the mass and decay width of the \CP-even (odd) eigenstate $D_{1(2)}$, respectively, and $\Gamma \equiv (\Gamma_1 + \Gamma_2)/2$ is the average decay width. Violation of the \CP symmetry in mixing leads to a non-unity value of the ratio $|q/p|$. When both \Dz and \Dzb mesons can decay to a common final state with amplitudes $\mathcal{A}$ and $\overline{\mathcal{A}}$, respectively, \CP violation can appear in the interference between mixing and decay amplitudes, and the quantity $\phi \equiv \arg[q\mathcal{A}/p\overline{\mathcal{A}}]$ differs from zero. 
 Violation of the \CP symmetry in mixing and in the interference between mixing and decay amplitudes are referred to as manifestations of {\it indirect} \CP violation. When \CP violation happens in the decay, $|\mathcal{A}/\overline{\mathcal{A}}|\neq 1$, this is referred to as {\it direct} \CP violation.

The importance of studying \Dz-\Dzb mixing is based on its nature of a flavor-changing neutral-current process, which is potentially sensitive to physics beyond the SM. The indirect search for new physics in flavor oscillations needs, however, precise predictions of the SM expectations to be compared with the measurements. Such predictions are very difficult to obtain in the charm sector. Theoretical methods like the quark-hadron duality and the heavy quark expansion work well in $b$-hadron physics, where the short-distance dynamics is dominant in the expansion of the energy release, $1/E_{\rm released}\sim 1/m_b$, being $m_b$ the $b$ quark mass. These methods can be applied for predicting $b$ hadron lifetimes and $B^0_{(s)}$-$\overline B^0_{(s)}$ mixing parameters. However, the charm quark mass ($m_c$) is not large enough compared to the energy release in charm decays, and long-distance dynamics are not negligible. Taking the $KKK$ intermediate state as an example, $E_{\rm released}\sim m_D-3m_K \sim \mathcal{O}(\Lambda_{QCD})$. Thus, quark-hadron quality must be handled carefully in the case of charm hadrons~\cite{Petrovtalk}. 

Except for the inclusive approach using the $1/m_c$ expansion just as in $B$ physics \cite{Bobrowski:2010xg}, the exclusive approach saturates the \Dz-\Dzb mixing correlators by hadronic states \cite{Falk:2001hx,Falk:2004wg,Cheng:2010rv,Jiang:2017zwr}. This is intriguing and intuitively plausible since the exclusive approach predicted that the values of $x$ and $y$ could reach the order of $1\%$, which is indeed experimentally observed \cite{Falk:2001hx,Falk:2004wg}. Flavor $SU(3)$-symmetry breaking plays a key role in the understanding of \Dz-\Dzb mixing, which would vanish in the limit of perfect $SU(3)$ symmetry, since $V_{cd}^*V_{ud}\approx-V_{cs}^*V_{us}$. 
The advantage of the exclusive approach is that the experimental uncertainties of hadronic branching fractions have been significantly improved nowadays. The current most precise prediction on the width difference is $y_{\rm PP+PV}=(0.21\pm0.07)\%$ \cite{Jiang:2017zwr}, considering the two-body contributions from the two pseudoscalar modes (PP) and the pseudoscalar-vector modes (PV),  using the Factorization-Assisted Topological amplitude approach, which includes the dominant $SU(3)$-breaking effects in the PP and PV modes \cite{Li:2012cfa,Li:2013xsa}. The result from two-body decays leads to expected values still smaller than the experimental data, implying additional contributions in \Dz-\Dzb mixing. 
It is proposed to consider the finite width effect as a new $SU(3)$-breaking effect~\cite{Petrovtalk}. The finite width effect of one-body intermediate states has been manifested in Ref.~\cite{Golowich:1998pz}. For 2-body contributions like pseudoscalar-vector, pseudoscalar-scalar, scalar-scalar, {\it etc.}, the vector, scalar and other excited states have finite widths and thus contribute to \Dz-\Dzb mixing as a new $SU(3)$-breaking effect. A Dalitz-plot analysis to understand the finite-width effects on \Dz-\Dzb mixing predictions is ongoing~\cite{Petrovtalk}.

A measurement of \Dz-\Dzb mixing and \CP violation can be obtained by comparing the ratio of $\Dz\to\Kp\pim$ and $\Dz\to\Km\pip$ decay rates, as a function of the \Dz decay time, with the corresponding ratio for the charge-conjugate processes. The flavour of the meson is determined either by the charge of the soft pion in $\Dstarp \to \Dz \pi^+_s$ decays or by the charge of the lepton (muon or electron) in semileptonic $\overline{B}$ decays, $\overline{B} \to \Dz l^- \nu_l X$. The soft pion or the lepton is then referred to as the tagging particle.  The decays dominated by a Cabibbo-favored amplitude, $\Dstarp \to \Dz(\to \Km \pip)\pi_s^+$ or $\overline{B} \to \Dz (\to \Km \pip) l^- \nu_l X$, are defined as ``right-sign'' (RS) decays. Wrong-sign (WS) decays, $\Dstarp \to \Dz(\to \Kp \pim)\pi_s^+$ or $\overline{B} \to \Dz (\to \Kp \pim) l^- \nu_l X$, are those due to the doubly Cabibbo-suppressed $\Dz \to \Kp \pim$ decay and the Cabibbo-favored $\Dzb\to \Kp \pim$ decay that follows a \Dz-\Dzb oscillation. The ratio of WS-to-RS rates as a function of the neutral \D-meson decay time $t$ can be approximated as
\begin{equation}
\label{eq:WSmix_rate}
    R(t) \approx R_D + \sqrt{R_D}\,y'\,\frac{t}{\tau} + \frac{x'^2 + y'^2}{4}\left( \frac{t}{\tau}\right)^2 \,,
\end{equation}
because of the small value of the mixing parameters. Here, $\tau$ is the average \Dz lifetime and $R_D$ is the ratio of suppressed-to-favored decay rates. The parameters $x'$ and $y'$ depend on the mixing parameters, $x' \equiv x\cos\delta + y\sin\delta$ and $y' \equiv y\cos\delta - x\sin\delta$, through the strong-phase difference $\delta$ between the suppressed and favored amplitudes. This phase was measured at the CLEO and BESIII experiments~\cite{Asner:2012xb,Ablikim:2014gvw}. 

A measurement of mixing and \CP violation parameters is reported by the LHCb collaboration using approximately $1.77\times10^8$ RS and $7.22\times10^5$ WS pion-tagged signal decays, reconstructed in a data sample corresponding to an integrated luminosity of 5.0\invfb from proton-proton ($pp$) collisions at 7, 8, and 13\tev center-of-mass energies~\cite{LHCb-ws-mixing}. Assuming \CP conservation, the mixing parameters are measured to be $x'^2 = (3.9 \pm 2.3 \pm 1.4) \times 10^{-5}$, $y' = (5.28 \pm 0.45 \pm 0.27) \times 10^{-3}$, and $R_D = (3.454 \pm 0.028 \pm 0.014) \times 10^{-3}$. 
Allowing for possible \CP violation, the decay-rate ratios of Eq.~\ref{eq:WSmix_rate} of mesons produced as \Dz and \Dzb are not the same. Indirect \CP violation generates differences in the value of the parameters $(x')^2$ and $y'$ for \Dz and $\Dzb$ mesons, which can constraints the value of $|q/p|$. If direct \CP violation occurs, the asymmetry $A_D \equiv (R_D^+ - R_D^- )/(R_D^+ + R_D^- )$ differs from zero, being $R_D^+$ and $R_D^-$ the ratio of suppressed-to-favored decay rates for \Dz and \Dzb mesons. Studying \Dz and \Dzb decays separately LHCb determines $1.00 < |q/p| < 1.35$ at the $68.3\%$ confidence level, and $A_D = (-0.1 \pm 8.1 \pm 4.2)  \times 10^{-3}$. These are the current most stringent bounds on the parameters $A_D$ and $|q/p|$ from a single measurement, and show no evidence for \CP violation. 

A sensitive probe of indirect \CP violation is given by the study of \Dz mesons decays into \CP eigenstates $h^+h^-$ ($h = K, \pi$). Being the time-integrated \CP asymmetries and the  mixing parameters known to be small, the decay-time-dependent \CP asymmetry of these decays can be approximated as
\begin{equation}
\label{eq:A_CP_t}
    A_{\CP}(t) =\frac{\Lambda(\Dz(t)\to h^+h^-) - \Lambda(\Dzb(t)\to h^+h^-)}{\Lambda(\Dz(t)\to h^+h^-) + \Lambda(\Dzb(t)\to h^+h^-)} \approx a_{\rm dir}^{h^+h^-} - A_\Gamma \frac{t}{\tau}\,,
\end{equation}
where $\Lambda(\Dz(t)\to h^+h^-)$ and $\Lambda(\Dzb(t)\to h^+h^-)$ are the decay-time-dependent rates of an initial \Dz or \Dzb decaying into $h^+h^-$ at decay time $t$, $a_{\rm dir}^{h^+h^-}$ is the asymmetry related to direct \CP violation and $A_\Gamma$ is the asymmetry between the \Dz and \Dzb effective decay widths,
\begin{equation}
    A_\Gamma \equiv \frac{\hat{\Gamma}_{\Dz \to h^+h^-} - \hat{\Gamma}_{\Dzb \to h^+h^-}}{\hat{\Gamma}_{\Dz \to h^+h^-} + \hat{\Gamma}_{\Dzb \to h^+h^-}}\,.
\end{equation}
Here, we have defined the effective decay width
$\hat{\Gamma}_{\Dz \to h^+h^-}$ as the inverse of the effective lifetime  $\int_0^\infty t\,\Lambda(\Dz(t)\to h^+h^-) {\rm d}t / \int_0^\infty \Lambda(\Dz(t)\to h^+h^-) {\rm d}t$. In fact, $A_{\Gamma}$ is the opposite of the indirect \CP asymmetry $a^{h^+ h^-}_{\rm ind}$.

Because of mixing, the effective decay width of decays to \CP-even final states $h^+h^-$  differs from the average decay width, $\Gamma=1/\tau$, measured in $\Dz \to \Km \pip$ decays. The quantity
\begin{equation}
    y_{\CP} \equiv \frac{\hat{\Gamma}_{\Dz \to h^+h^-} + \hat{\Gamma}_{\Dzb \to h^+h^-}}{2\Gamma} - 1
\end{equation}
is equal to $y$ if \CP symmetry is conserved. Neglecting the $\mathcal{O}(10^{-3})$ difference between the phases of the $\Dz \to \Kp\Km$ and $\Dz \to \pip\pim$ decay amplitudes, $A_\Gamma$ and $y_{\CP}$ are independent of the $h^+h^-$ final state, and $\phi \approx \arg(q/p) $. In the limit of small \CP violation in mixing ($|q/p| \approx 1$), $A_\Gamma$ can be approximated as $-x \sin\phi$ and its measurement translates into a constraint of the mixing phase $\phi$, if the value of $x$ is determined elsewhere. For $y_\CP$ instead, $y_\CP\approx y \cos\phi$. 

In a data sample corresponding to an integrated luminosity of 3\invfb, LHCb selects about 9.6 and 3.0 million of pion-tagged $\Dz \to \Kp\Km$ and $\Dz \to \pip\pim$ decays, respectively, to measure $A_\Gamma(\Kp\Km) = (-0.30 \pm 0.32 \pm 0.10) \times 10^{-3}$ and $A_\Gamma(\pip\pim) = (0.46 \pm 0.58 \pm 0.12)\times 10^{-3}$~\cite{LHCb-AGamma}. The results are consistent and show no evidence of \CP violation; they are combined to yield a single value of $A_\Gamma =(-0.13\pm0.28\pm0.10)\times10^{-3}$. Combination with a smaller, independent muon-tagged sample~\cite{LHCb-AGamma-muon}, gives $A_\Gamma = (-0.29 \pm 0.28)\times10^{-3}$.  The measurements of $A_\Gamma$ reported by LHCb are the most precise to date. They are consistent with results from Belle obtained from the analysis of the full data set collected by the experiment, corresponding to an integrated luminosity of 976\invfb, in which $A_\Gamma$ and $y_{\CP}$ are measured simultaneously with $\tau$, using  $242\times10^3$ $\Dz \to \Kp\Km$, $114\times10^3$ $\Dz \to \pip\pim$ and $2.6\times10^6$ $\Dz \to \Km\pip$ decays, all from $\Dstarp \to \Dz \pi_s^+$ decays~\cite{Belle-AGamma}. The values found by Belle are $y_{\CP} = (1.11 \pm 0.22 \pm 0.09)\%$, $A_\Gamma =(-0.3\pm2.0\pm0.7)\times10^{-3}$ and $\tau=(408.46\pm 0.54)$\,fs.

The precision on $y_{\CP}$ is greatly improved by a brand new results presented by LHCb at the workshop for the first time~\cite{LHCb-yCP}. In a data sample corresponding to an integrated luminosity of 3\invfb, $880\times10^3$ $\Dz \to \Kp\Km$, $310\times10^3$ $\Dz \to \pip\pim$ and $4.6\times10^6$ $\Dz \to \Km\pip$ decays, are reconstructed from semimuonic decays of $\overline{B}$ mesons. The difference between the widths of \Dz decays to $h^+h^-$ and $\Km\pip$ final states, $\Delta\Gamma \equiv \hat{\Gamma}-\Gamma$, is measured from a fit to the ratio between $\Dz \to h^+h^-$ and $\Dz \to \Km\pip$ signal yields, corrected for the selection efficiency, as a function of the \Dz decay time. The parameter $y_{\CP}$ is then calculated from the measured value of $\Delta\Gamma$ and the known value of $\Gamma$ as $y_{CP} = \Delta\Gamma/\Gamma$. The results are $y_{\CP}(\Kp\Km) =(0.63\pm0.15\pm0.11)\%$, and $y_{\CP}  =(0.38\pm0.28\pm0.15)\%$, consistent between each other. The value of $y_{\CP}(\Kp\Km)$ is the most precise to date from a single experiment. The two measurements are combined and yield $y_{\CP} = (0.57 \pm 0.13 \pm 0.09)\%$, which is as precise as the current world average value, $(0.84 \pm 0.16)\%$~\cite{hflav}. The result is also consistent with the known value of the mixing parameter $y =(0.67^{+0.06}_{-0.13})\%$~\cite{hflav}, showing no evidence for \CP violation in mixing. A measurement of $y_{\CP}$ in correlated $D^0\bar{D}^0$ is reported also by the BESIII experiment~\cite{Mix:BESIII:2015}, through a comparison of the combined branching fractions of several \CP-even ($\Kp\Km$, $\pip\pim$, $\KS\piz\piz$) and \CP-odd ($\KS\piz$, $\KS\omega$, $\KS\eta$) final states, although the sensitivity is low, around $1.5\%$. BESIII is also updating the study by including $K_L^0 \piz (\piz)$.

Mixing and \CP violation are also studied through multibody decays. The golden decay mode in this case is $\Dz \to \KS \pip\pim$, which final state is a complex assembly of different resonances including flavour and \CP eigenstates. Sensitivity to mixing and \CP violation can be enhanced by multiple interfering amplitudes with strong phases continuously varying across phase-space. Two main methods have been proposed to study these decays: the model-independent and the model-dependent method.  A new model-independent method based on a Fourier analysis was also proposed for a measurement of the CKM angle $\gamma$~\cite{Poluektov:2017zxp}, but not yet employed in a measurement of charm mixing. 

Model-dependent analyses measure effective lifetime of individual resonances, and the choice of the model to describe the Dalitz plot represents  irreducible systematic uncertainties. This method has been used at $B$ factories~\cite{Babar-KSpipi,Belle-KSpipi}. Model-independent measurements study  the decay-time evolution of \Dz mesons in bins of the Dalitz plot, according to similar strong phase differences between the interfering amplitudes. This type of analysis is used by the LHCb experiment~\cite{LHCb-KSpipi}. The major limitation comes from the external input required on the values of the strong phase differences of the interfering resonances. These are measured at the  CLEO-c and BESIII experiments, exploiting the production of correlated pairs of charm mesons at threshold. A preliminary (since 2014) result from BESIII~\cite{BESIII-prel-KSpipi} provides an improvement of about 40\% on the uncertainty of the measurement by CLEO-c~\cite{CLEO-KSpipi}, currently the one exploited in the LHCb analysis; however, the BESIII analysis is still in progress, and no update (especially on the missing systematic uncertainties) has been presented at the workshop. Such a measurement would also be important input in the measurement of the CKM angle $\gamma$ using the so-called GGSZ method~\cite{GGSZ-method}. 
The results on the mixing parameters $x$ and $y$ obtained with $\Dz\to\KS\pip\pim$ decays from different experiments are reported in Table~\ref{tab:x_y_KSpipi}. Updates from LHCb are expected soon. 
\begin{table}[t]
\caption{Mixing parameters $x$ and $y$ obtained with $\Dz\to\KS\pip\pim$ decays from different experiments. The result from BaBar is a combination of $\Dz\to\KS\pip\pim$ and $\Dz\to\KS\Kp\Km$ decays.}\label{tab:x_y_KSpipi}
\centering
\begin{tabular}{lcc}
\toprule
             & $x$ $[\%]$ & $y$ $[\%]$ \\
\midrule
LHCb~\cite{LHCb-KSpipi}   &  $-0.86 \pm 0.53 \pm 0.17$ &  $-0.03 \pm 0.46 \pm 0.13$\\
Belle~\cite{Belle-KSpipi} &  $0.56 \pm 0.19^{+0.07}_{-0.13}$ & $0.30 \pm 0.15^{+0.05}_{-0.08}$\\
BaBar~\cite{Babar-KSpipi}  & $0.16 \pm 0.23 \pm 0.14$ &  $0.57 \pm 0.20 \pm 0.15$\\
\bottomrule
\end{tabular}
\end{table}

These results can be compared with those obtained by BaBar from a model-dependent analysis of $138 \times 10^3$ $\Dz \to \pip\pim\piz$ decays, originating from $\Dstarp \to \Dz \pi^+_s$ decays, using a sample corresponding to an integrated luminosity of 468\invfb, $x=(1.5\pm1.2\pm0.6)\%$ and $y=(0.2\pm0.9\pm0.5)\%$~\cite{Babar-pipipiz}. 
At LHCb, mixing is also observed in a decay-time-dependent analysis of the WS-to-RS ratio of $\Dz\to \Kp\pim\pip\pim$ decay rates~\cite{LHCb-K3pi}, obtaining $x=(0.41\pm0.17)\%$ and $y=(0.67\pm0.80)\%$. This analysis is sensitive to the phase-space averaged ratio of doubly Cabibbo-suppressed to Cabibbo-favoured amplitudes $r_D^{K3\pi}$, measured to be $r_D^{K3\pi} = (5.67\pm0.12)\times10^{-2}$, and the product of the coherence factor $R_D^{K3\pi}$ and a charm mixing parameter $y'_{k3\pi}$, found to be $R_D^{K3\pi} y' = (0.3 \pm 1.8) \times 10^{-3}$. This provides useful input for determinations of the CKM angle $\gamma$ in $B^+ \to \Dz(\to \Km \pip \pim\pip) \Kp$ decays, for which the parameters $r_D^{K3\pi}$, $R_D^{K3\pi}$ and the strong phase $\delta_D^{K3\pi}$ are required. A combination of the LHCb results with CLEO-c data significantly improves the precision on these inputs~\cite{CLEO-K3pi}.

%% file: directCPV.tex
\section{Direct \boldmath{\CP} violation}
\label{sect:direct_CPV}
Direct \CP violation in charm decays is difficult to be predicted precisely in the SM due to the relatively large non-perturbative effects. It is nonetheless expected to be small due to the GIM mechanism with $m_{d,s,b}\ll m_W$. Thus, \CP violation in charm decays is of good use as a null test of the SM, as measurements of large \CP asymmetries (generally larger than $\mathcal{O}(1\%)$) would be indication of new physics. In addition, contrary to the case of the $K$ and $B$ systems, the charm system is unique for studying \CP violation in the up sector of quarks. This is helpful to distinguish new physics models with different quark structures. The recent flavor anomalies~\cite{Lees:2012xj,Huschle:2015rga,Aaij:2015yra} indicate a possible large non-SM \CP phase, which may increase \CP violation in charm decays \cite{Sonitalk}. For example, the $R(D^{(*)})$ anomaly points to a non-SM structure of the $cb$ quarks coupling. Concerning \CP violation, the $cb$ coupling has no complex phase in the SM, whereas it might present one in a new physics scenario. This might manifest in measurements of direct \CP violation in charm decays, no matter what the charm-penguin amplitude considered. 

The decay-time-dependent \CP asymmetry, $A_{\CP}(t)$, for \Dz mesons decaying to a \CP eigenstate $h^+h^-$ is given by Eq.~\ref{eq:A_CP_t}. 
In a sample of reconstructed $\Dz \to h^+h^-$ decays, the decay-time-integrated asymmetry, $A_{\CP}$, depends upon the reconstruction efficiency as a function of the decay time, and can be written as $A_{\CP} \approx  a_{\rm dir}^{h^+h^-} ( 1 + \langle t\rangle/\tau\,y_{\CP} ) - \langle t\rangle/\tau\,A_\Gamma$, where $\langle t\rangle$ denotes the mean of the reconstructed decay-time distribution of the $\Dz\to h^+h^-$ decays. What is experimentally accessible is the {\it raw} asymmetry of the number of candidate, $N$, counted as decay of a \Dz or a \Dzb mesons,
\begin{equation}
    A_{\rm raw} \equiv \frac{N(\Dz\to h^+h^-) - N(\Dzb\to h^+h^-)}{N(\Dz\to h^+h^-) + N(\Dzb\to h^+h^-)}\,.
\end{equation}
The raw asymmetry can be written, up to $\mathcal{O}(10^{-6})$, as $A_{\rm raw} \approx A_{\CP} + A_f + A_{\rm tag} + A_{\rm prod}$, where $A_{f}$ and $A_{\rm tag}$ are the asymmetries in the reconstruction efficiency of the \Dz final states and of the tagging particle, and $A_{\rm prod}$ is the production asymmetry for either \Dstarp or $\overline{B}$ mesons (relevant only at hadron colliders). For self-conjugate final states, $A_f=0$. To a good approximation, $A_{\rm tag}$ and $A_{\rm prod}$ are independent of the final state in any given kinematic region. Thus, the difference of raw asymmetry between $\Dz\to \Kp\Km$ and $\Dz\to \pip\pim$ decays give access to 
\begin{equation}
    \Delta A_{\CP} \equiv A_{\CP}^{\Kp\Km} - A_{\CP}^{\pip\pim} \approx \Delta a_{\rm dir} \left( 1 + \frac{\overline{\langle t\rangle}}{\tau}y_{\CP} \right) - \frac{\Delta\langle t\rangle}{\tau}A_\Gamma\,,
\end{equation}
where $\overline{\langle t\rangle}$ is the arithmetic mean of $\langle t^{\Kp\Km}\rangle$ and $\langle t^{\pip\pim}\rangle$.  This equation holds up to terms of $\mathcal{O}(10^{-6})$. In the limit of exact U-spin symmetry, $\Delta a_{\rm dir}\equiv a_{\rm dir}^{\Kp\Km} - a_{\rm dir}^{\pip-\pim} \approx 2 a_{\rm dir}^{\Kp\Km}$, since $a_{\rm dir}^{h^+h^-}$ is expected to be equal in magnitude and opposite in sign for $\Kp\Km$ and $\pip\pim$ final states~\cite{Uspin_DACP}, although U-spin breaking effects could be also present~\cite{Uspin_break_DACP}. 

To take into account an imperfect cancellation of detection and production asymmetries due to the difference in the kinematic properties of the two decay modes, a weighting procedure in bins of kinematic distributions is usually employed to equalize the kinematic of $\Kp\Km$ and $\pip\pim$ final state. With this method, LHCb obtains the most precise measurement of $\Delta A_{\CP}$ to date, $\Delta A_{\CP} = (-0.10\pm0.08\pm0.03)\%$~\cite{LHCb-DACP-pion, LHCb-DACP-muon}, from a combination of the results from the pion- and muon-tagged samples. This measurement exploits only $1/3$ of the currently available data: the analysis of the full data set is ongoing, and the statistical precision is expected to reach an unprecedented value, around $0.03\%$. 

The direct \CP asymmetry in the $\Dz \to \KS\KS$ decay is also interesting to test the SM prediction of $A_{\CP}\leq 1.1\%$, or to signal new physics if data exceeds it \cite{Nierste:2015zra}. Both $B$-factories and LHCb presented a measurements of  \CP asymmetry in this decay. At LHCb, the analysis exploits again the combination of the raw asymmetries measured in different decays to cancel spurious asymmetries. The $\Dz \to \Kp\Km$ decay is used as the calibration channel to cancel the production and tagging asymmetry, owing on the precise measurement of $A_{\CP}^{\Kp\Km}$~\cite{LHCb-ACP-KK}. Thus, $A_{\CP}^{\KS\KS}$ is obtained from a measurement of $A_{\CP}^{\KS\KS} - A_{\CP}^{\Kp\Km}$. The result presented at the workshop exploits the data collected by LHCb in 2015 and 2016, corresponding to an integrated luminosity of 2\invfb, and is $A_{\CP}^{\KS\KS} =(4.2\pm3.4\pm1.0)\%$~\cite{LHCb-KSKS}. Combining this with a previous result from an independent sample~\cite{LHCb-KSKS-1}, LHCb obtains $A_{\CP}^{\KS\KS} =(2.0\pm2.9\pm1.0)\%$, which doesn't hint to \CP violation. This is in agreement with a more precise result presented by Belle, $A_{\CP}^{\KS\KS} =(-0.02 \pm 1.53 \pm 0.17)\%$, obtained from a sample corresponding to an integrated luminosity of 921\invfb~\cite{Belle-KSKS}. In the same analysis, Belle provides also a measurement of the branching fraction of this decay, $\mathcal{B}(\Dz \to \KS\KS) = (1.321 \pm 0.023 \pm 0.057) \times 10^{-4}$.

Thanks to the large data samples accumulated, the study of \CP violation of rare charm decays is also possible. A first measurement of \CP asymmetries in the rare $\Dz\to h^+h^-\mup\mum$ decays is presented by the LHCb collaboration~\cite{LHCb-rareD}, reconstructing about $1000$ $\Dz\to \pip\pim\mup\mum$ and $100$ $\Dz\to\Kp\Km\mup\mum$ decays in a data set corresponding to an integrated luminosity of 5\invfb. In these decays, \CP asymmetry is expected to be below $5\times10^{-3}$ in the standard model, but enhanced up to about 1\% if new physics is present. Forward-backward and angular asymmetries are also measured, but they were discussed at the workshop in more details by the WG3~\cite{WG3summary}.  Here, we focus on the \CP asymmetry, which is measured in a similar fashion as $A_{\CP}^{\KS\KS}$, {\it i.e.} by using the $\Dz \to \Kp\Km$ decays as a calibration channel to cancel the production and tagging asymmetry, and measuring  $A_{\CP}^{h^+h^-\mup\mum} - A_{\CP}^{\Kp\Km}$. Using the precise known value of $A_{\CP}^{\Kp\Km}$, LHCb determines $A_{\CP}^{\Kp\Km\mup\mum}=(0 \pm 11\pm 2)\%$ and $A_{\CP}^{\pip\pim\mup\mum}=(4.9\pm3.8\pm0.7)\%$, both in agreement with \CP conservation in these decays. 

Considering decays of the $\Dp$ meson, the BESIII collaboration presents measurement of \CP asymmetries in singly Cabibbo-suppressed decays $\Dp \to \KS \Kp$, $\Dp \to \KS \Kp \piz$, $\Dp \to \KL \Kp$ and $\Dp \to \KL \Kp \piz$, obtaining the results: 
$A_{\CP}(\KS \Kp)=(-1.5\pm2.8\pm1.6)\%$,
$A_{\CP}(\KS \Kp \piz)=(1.4\pm4.0\pm2.4)\%$,
$A_{\CP}(\KL \Kp)=(-3.0\pm3.2\pm1.2)\%$, and
$A_{\CP}(\KL \Kp \piz)=(-0.9\pm4.1\pm1.6)\%$~\cite{Weidenkaff}. These do not indicate \CP violation.
 Note that the last three measurements are provided for the first time. 
The decay channel $D^+\to\pi^+\pi^0$ is another excellent candidates to probe \CP violation in charm decays for searching for new physics~\cite{DCPV:Grossman:2012, DCPV:Buccella:1993}.
The Belle collaboration reports a measurement which exploits a sample of  about $6.6\times 10^{3}$ pion-tagged and $101\times 10^{3}$ untagged $D\to\pi^+\pi^0$ decays. They measures $A_{\CP}(D^+ \to \pi^+ \pi^0) =(2.31\pm1.24\pm0.23)\%$ and $A_{\CP}(\Delta I = 1/2)=(-2.2\pm2.7)\%$~\cite{DCPV:Belle:2018}. At Belle II, the precision is expected to improve  to reach uncertainties of about $0.18\%$ (statistical) and $0.20\%$ (systematic).

Results on radiative charm decays $D^0 \to V \gamma$, where $V$ is a vector meson, are also reported. Measurements of branching fractions of these decays are useful test for QCD-based models, since their amplitudes are dominated by non-perturbative long range dynamics.  These decays are also sensitive to new physics searches via measurement of \CP asymmetries: it has been suggested that those asymmetries can rise to several percent in contrast to $\mathcal{O}(10^{-3})$ SM expectation in several non-SM scenarios~\cite{DCPV:Isidori:2012,DCPV:Lyon:2012,deBoer:2015boa}. 
The Belle collaboration performed first measurements of \CP violation in $D^0 \to V\gamma$
decays using 943\invfb of data~\cite{DCPV:Belle:2017}, obtaining the results reported in Tab.~\ref{tab:radiative_results}. They
are consistent with zero \CP asymmetry in any of the $D^0 \to V \gamma$ decay mode. 

Study of photon polarization in these decays is also a sensitive probe to non-SM physics. In the inclusive decay of $c\to u\gamma$, the photon polarization is proportional to $r=C'_7/C_7$, which is almost a null test of the SM. In the exclusive processes, the photon polarization suffers the pollution of long-distance contributions, $r^{\rm SM}=\mathcal{O}(\Lambda_{\rm QCD}/m_c)$ \cite{dBoertalk}. It is proposed to use data and $U$-spin symmetry to relate $D^0\to \rho^0\gamma$ and $D^0\to \overline K^{*0}\gamma$ decays \cite{deBoer:2018zhz}. For decays of the type 
$D\to A(\to P_1P_2P_3)\gamma$, where $A$ is an axial-vector meson, the photon polarization can be determined by the up-down asymmetry~\cite{Adolph:2018hde}. The charm decays have an advantage compared to $B$ decays: the phase spaces are suppressed for higher resonances \cite{deBoer:2018zhz}. The angular asymmetry in charm baryon decays, such as $\Lambda_c^+\to p \gamma$, is also sensitive to the chirality-flipped contributions \cite{deBoer:2017que}.

\begin{table}[t]
\caption{Measurements of branching fractions and \CP asymmetries in radiative charm decays presented by the Belle collaboration~\cite{DCPV:Belle:2017}.}\label{tab:radiative_results}
\centering
\begin{tabular}{lcc}
\toprule
     Decay        &  $\mathcal{B}$ $[10^{-5}]$ & $A_{\CP}$  $[\%]$ \\
\midrule
$D^0 \to \phi \gamma$   &  $2.76\pm 0.19\pm0.10$ &  $-9.4\pm6.6\pm0.1$\\
$D^0 \to \bar{K}^{*0} \gamma$ &  $46.6\pm 2.1\pm2.1$ & $-0.3\pm2.0\pm 0.1$\\
$D^0 \to \rho^0 \gamma$  & $1.77\pm 0.30\pm0.07$ &  $5.6\pm15.2\pm0.6$\\
\bottomrule
\end{tabular}
\end{table}

Since \CP asymmetries are  proportional to the ratio of penguin and tree amplitudes, $A_{\CP}\propto |P/T|$, to seek out a process with enhanced \CP violation, it is suggested to look at decays where the penguin amplitude is enhanced and the tree amplitude is suppressed as much as possible~\cite{Sonitalk}. The strategy is, i) to avoid $W^+\to u\bar d$ or $u\bar s$ making charged vector states, such as $\rho^\pm$ or $K^{*\pm}$, and ii) to go for color suppressed or Zweig suppressed final states from tree amplitudes. A Cabibbo-suppressed decay is automatically forced by the tree-penguin interference. Possible decay modes are listed in Ref.~\cite{Atwood:2012ac}.

All of the above mentioned \CP asymmetries arise from the interference between tree and penguin amplitudes. A basic shortcoming is the unknown penguin dynamics. We thus lose the power of prediction on \CP violation in charm induced by the tree-penguin interference. On the contrary, the tree amplitudes are better known with non-perturbative hadronic matrix elements extracted from measurements of branching fractions~\cite{Cheng:2010ry,Li:2012cfa,Li:2013xsa,Muller:2015lua}. 
It deserves to study the tree-tree interference-induced \CP violation, {\it i.e.} the interference between Cabibbo-favored (CF) and doubly Cabibbo-suppressed (DCS) amplitudes,  {\it e.g.} in $D^+\to\pi^+K_S^0$ decay~\cite{Yu:2017oky}. In literature, it was always postulated that, subtracting the \CP violation in kaon mixing, data would reveal a direct \CP asymmetry in such charm decays. However, this is not correct when considering the neutral kaon reconstructed by two charged pions~\cite{Yu:2017oky}. To consider with the well known \CP violation in kaon mixing and direct \CP asymmetry, a new \CP-violating effect is found for the first time, arising from the interference between the mother decay and the daughter mixing. This is more complicated than the ordinary mixing-induced \CP asymmetry in which both decay and mixing occur in the mother particle, for example the $B^0\to \pi^+\pi^-$ decay. The key point is that we have to consider all the amplitudes of the decay chain, say, the CF and DCS amplitudes of the charm decaying into neutral kaons, and the neutral kaons reconstructed by two charged pions. 
In the previous studies, either the DCS amplitude was neglected \cite{Grossman:2011zk}, or the decay chain of neutral kaons decaying into two charged pions was not considered \cite{Bigi:1994aw,Xing:1995jg}. Therefore, the total \CP asymmetries in the charm decays into neutral kaons is actually 
\begin{align}
A_{\CP}(t)=A_{\CP}^{\overline K^0}(t) + A_{\CP}^{\rm dir}(t) + A_{\CP}^{\rm int}(t)\,,
\end{align}
where the first two terms are the well known \CP asymmetries in kaon mixing and the direct \CP violation in charm decays, respectively,  and the third term is the new \CP-violating effects induced by the interference between charm decays and kaon mixing. Using the factorization-assisted topological-amplitude (FAT) approach~\cite{Li:2012cfa,Wang:2017ksn}, 
it is found that the new \CP-violating effect can be as large as the order of $10^{-3}$, thus is accessible at the LHCb and Belle II experiments in the near future. It is proposed to measure the new \CP-violating effect via
\begin{align}
\Delta A_{\CP} = A_{\CP}(D^+\to \pi^+K_S^0) - A_{\CP}(D_s^+\to K^+K_S^0)\,,
\end{align}
where the dominated \CP violation from kaon mixing is cancelled, and the direct \CP asymmetry in charm decays is negligible. 
Finally, the new \CP-violating effect has to be considered in the studies of new physics using the direct \CP asymmetry of charm. 

%% file: Todd.tex
\section{\boldmath{\CP}-violating triple product asymmetries}
\label{sect:TP_asym}

Time-reversal asymmetry is sensitive to \CP violation via the $\CP T$ theorem ($T$ violation implies $\CP$ violation under the assumption of $\CP T$ invariance.).  Measurements of $T$-odd asymmetries are a clean way to search for \CP violation in the charm sector. Analysis of four-body charm decays allow to probe \CP violation in different phase space regions, where sensitivity can be enhanced due to several interfering amplitudes with different relative strong phases. In four body $D$ decays, $D \to a b c d$, one can use triple products of final-state particles momenta (in the $D$ frame),
 $C_T \equiv \vec{p}_c\cdot (\vec{p}_a \times \vec{p}_b)$ and 
$\overline{C}_T \equiv \vec{p}_{\overline{c}}\cdot (\vec{p}_{\overline{a}} \times \vec{p}_{\overline{b}}) $
 to construct two $T$-odd observables,
 \begin{equation}
    A_T \equiv \frac{\Gamma(C_T>0)- \Gamma(C_T<0)}{\Gamma(C_T>0)+ \Gamma(C_T<0)},\mbox{ and }
    \overline{A}_T \equiv \frac{\Gamma(-\overline{C}_T>0)- \Gamma(-\overline{C}_T<0)}{\Gamma(-\overline{C}_T>0)+ \Gamma(-\overline{C}_T<0)}\,,
 \end{equation}
 where $\Gamma_D$ ($\Gamma_{\overline{D}}$) is the decay width of $D$ ($\overline{D}$) decays to $abcd$ ($\overline{abcd}$) in the $C_T$ ($\overline{C}_T$) range. However $A_T$ ($\overline{A}_T$) can be non-zero due to final state interaction (FSI) even without \CP violation. In order to cancel FSI, one can create genuine $\CP$-violating asymmetry (insensitive also to $D$ production asymmetry and charged-particle reconstruction asymmetries) as
 \begin{equation}
     \atodd =\frac{1}{2}(A_T - \overline{A}_T)\,.
  \end{equation}
  A triple product is even (odd) under $C$ ($P$, $T$, and $\CP$). One can construct a number of asymmetries that can be computed by integrating over positive and negative values of the triple product~\cite{Bevans}.
 
 First measurement of \atodd in decays $D^0\to K^+K^-\pip \pim$ and $D^+\to \KS K^+\pip \pim$ was done by FOCUS~\cite{Todd:Focus:2005}. The precision of these measurements was then improved by BaBar, and the study of the decay $D_S^+ \to K_S^0 K^+\pip\pim$ was added~\cite{Todd:BaBar:2010b}. Asymmetries consistent with zero were measured.
 The LHCb collaboration also measured  $\atodd(D^0\to K^+K^-\pip\pim)$, obtaining the precise result $(0.18\pm0.29\pm0.04)\%$, consistent with zero~\cite{Todd:LHCb:2014}. For the first time, LHCb also presented $\atodd$ in different ranges of phase space and in bins of $D^0$ decay time. Local asymmetries up to 30\% are seen in bins of phase space. The Belle experiment reported the first measurement of the $\atodd(D^0 \to \KS \pip \pim \piz)$ to be $(-0.28 \pm 1.38^{+0.23}_{-0.76}) \times 10^{-3}$, consistent with no $\CP$ violation~\cite{Todd:Belle:2017}, one of the most precise measurement. They also perform $\atodd$ measurements in different region of $D^0 \to K_S^0  \pip \pim \piz$ phase space, with no evidence of $\CP$ violation.
 
In this workshop, Belle reported their new result on the four body decay of $D^0 \to K^+ K^- \pip \pim$, measuring a set of five kinematically independent $\CP$ asymmetries, of which four asymmetries are measured for the first time~\cite{Todd:Belle:2018}. They are reported in Tab.~\ref{tab:todd}, where in the notation $\theta_1$ ($\theta_2$) is the helicity angle of $K^+K^-$ ($\pip\pim$) systems against positive charged particles, and $\phi$ is the angle between decay planes of these systems. No $\CP$ violation is observed, however, these new kinematic asymmetries further constrain new physics models for the first time.

\begin{table}[t]
\caption{Measurement of \CP-violating kinematic asymmetries measured by Belle~\cite{Todd:Belle:2018}.}
\label{tab:todd}
\centering
\begin{tabular}{lc}
\toprule
   Asymmetry          & Result $[\times10^{-3}]$\\
\midrule
$a_{\cos\phi}$ & $3.4\pm3.6\pm0.6$\\
$a_{\sin\phi}$ & $5.2\pm3.7\pm0.7$ \\
$a_{\sin2\phi}$ & $3.9\pm3.6\pm0.7$\\
$a_{\cos\theta_1\cos\theta_2\cos\phi}$ & $-0.2\pm3.6\pm0.7$\\
$a_{\cos\theta_1\cos\theta_2\sin\phi}$ & $0.2\pm3.7\pm0.7$ \\
\bottomrule
\end{tabular}
\end{table}

%% file: baryons.tex
\section{Results on charm baryons}
\label{sect:baryons}

Violation of the \CP symmetry in charm baryons decays is expected to be at the same level as in charm mesons decays.  
A first measurement of \CP violation in a 3-body decay of the \Lc baryons is reported by the LHCb collaboration, by studying the Cabibbo-suppressed decays $\Lc \to p h^+h^-$ originating from semimuonic \Lb decays~\cite{LHCb-ACP-Lc}. Similar to the $\Delta A_{\CP}$ measurement of $\Dz\to h^+h^-$ decays described in Sect.~\ref{sect:direct_CPV}, the analysis measures the difference in the raw asymmetry between the $\Lc \to p \Kp\Km$ and $\Lc \to p \pip\pim$ to cancel spurious asymmetries, and access the \CP-asymmetry difference between the two decay modes. The spurious asymmetries are those due to asymmetries in the reconstruction efficiency of the proton in the final state, of the muon from the \Lb decays, and of the \Lb production. To allow for their precise suppression, kinematic distributions of the \Lb, the proton and the muon of the $\Lc \to p \pip\pim$ sample are weighted to match those of the $\Lc \to p \Kp\Km$ sample. Since $\Lc \to p h^+h^-$ dynamics are described by a 5-dimensional phase space, along which \CP violation effects may vary, efficiency corrections are considered across the full phase space. From about $25\times10^3$ $\Lc \to p \Kp\Km$ and $161\times10^3$ $\Lc \to p \pip\pim$ decays, the LHCb collaboration obtains $A_{\CP}^{p \Kp\Km} - A_{\CP}^{p \pip\pim} = (0.30 \pm 0.91 \pm 0.61)\%$, finding no signal of \CP violation. 

 Owing on the large cross-section for charm production in $pp$ collisions, the LHCb experiment can collect large data set of charm baryons decays and open new avenues in this sector. Indeed, interesting results which go behind measurements of \CP violation were reported at the workshop. A breakthrough of double-charm baryon physics happened during the past two years. In 2017 it was firstly pointed out in \cite{Yu:2017zst} that $\Xi_{cc}^{++}\to \Lambda_c^+K^-\pi^+\pi^+$ and $\Xi_c^+\pi^+$ are the most favorable modes to search for doubly heavy flavor baryons, with branching fractions of $\mathcal{O}(10\%)$. Soon afterwards, the LHCb collaboration discovered a first doubly charmed baryon, $\Xi_{cc}^{++}$, via the final states of $\Lambda_c^+K^-\pi^+\pi^+$ \cite{LHCb-Xicc-obs}. The LHCb collaboration keeps improving the understanding of the newly observed \Xiccpp baryon in 2018, by reporting the observation of $\Xiccpp \to \Xicp \pip$~\cite{LHCb-Xicc-obs-2}, with $\Xicp\to p\Km\pip$, with a significance of $5.9$ standard deviations. A new measurement of the \Xiccpp mass is obtained from this decay mode, $m(\Xiccpp) = 3620.6 \pm 1.5 \pm 0.5 \mevcc$, and the combination of this result with the measurement in Ref.~\cite{LHCb-Xicc-obs} yields to $m(\Xiccpp) = 3621.24 \pm 0.65 \pm 0.31 \mevcc$. The analysis leads also to the measurement of the following combination of branching fractions, $\mathcal{B}(\Xiccpp \to \Xicp \pip)\mathcal{B}(\Xicp\to p\Km\pip)/\mathcal{B}(\Xiccpp\to \Lc\Km \pip\pip)\mathcal{B}(\Lc\to p\Km\pip) = 0.035 \pm 0.009 \pm 0.003$, which is consistent with the prediction in \cite{Yu:2017zst,Wang:2017mqp}. 

Another important result on the \Xiccpp baryon is the measurement of its lifetime~\cite{LHCb-Xicc-life}. The decay mode employed is the same used for the first observation, $\Xiccpp\to \Lc(\to p\Km\pip)\Km \pip\pip$.  A correction for the selection efficiency of the reconstructed decay-time distribution of the \Xiccpp decays is extracted from simulation. To suppress biases on this correction due to potential mismodeling of the simulation, the efficiency-corrected distribution is normalized to the efficiency-corrected distribution of $\Lb \to \Lc(\to p\Km\pip) \pim \pip\pim$ decays, reconstructed in the same data set with a similar selection. From the ratio of the two decay-time distribution, taking in input the precisely known value of the \Lb lifetime, the \Xiccpp lifetime is found to be $256^{+22}_{-20}\pm14$\,fs.

A new lifetime measurement is also reported by LHCb for the \Omegac baryon~\cite{LHCb-Omegac-life}. This baryon is reconstructed in the decay $\Omegac\to p \Km\Km\pip$, where the \Omegac originates from semileptonic decays, $\Omegab \to \Omegac \mum \nu_\mu X$. Collecting a signal yields of about 980 $\Omegac$ decays, {\it i.e.} ten times higher than any previous measurement, the lifetime is measurement using the $\Dp \to \Km \pip\pip$ decay, reconstructed from $B \to \Dp \mum \nu_mu X$ decays, as a normalization channel. This allows to suppress potential biases on the modeling of the decay-time-dependent efficiency (with a similar strategy adopted for the measurement of the \Xiccpp lifetime just reported). The value of the \Omegac lifetime found by LHCb is $268 \pm 24 \pm 10$\,fs, in large disagreement with the current world average, $69 \pm 12$\,fs~\cite{pdg}. This is also in contrast with the hierarchy $\tau(\Xicp)>\tau(\Lc)>\tau(\Xicz)>\tau(\Omegac)$, as this measurement leads to a value smaller than the \Xicp lifetime, but larger than that of the \Lc baryon. 

%% file: prospects.tex
\section{Belle 2 and LHCb upgrades}
\label{sect:prospects}
In 2018, Belle~II successfully completed the ``Phase II" commissioning and accumulated 472\invpb of data. All subdetectors except for the full vertex detector were inserted and operating. Initial plots of reconstructed charm decays were shown~\cite{casarosa,kumar}, suggesting the readiness of the experiment to enter the game for charm physics. Belle~II, with an ability to accumulate 50 times more data in comparison to Belle, is expected to play an important role in the next future. Due to the new vertex detector, which should be operational in 2019, Belle~II can exploit a factor of two of improvement with respect to Belle on the resolution of the track impact parameter. This allows to have a decay-time resolution in $D \to hh$ decays of about $0.14$\,ps, which is roughly two times better than that at Belle and BaBar~\cite{BelleIIBook}. It is also reported a good resolution, similar to that of the Belle experiment, for the $D$ mass in decays into final states with neutral particles  ({\it e.g.} $\pi^0$, $\eta$). This is important as Belle~II is expected to provide a reach program of measurement in decay channels with neutrals, which are more difficult to be reconstructed by LHCb. Thank also to the clean environment, Belle~II is expected to measure mixing via interference in modes such as $D^0 \to K^- \pi^+ \piz$ and achieve sensitivity of  about 0.06\% (0.05\%)  for  $\sigma_{x"}$ ($\sigma_{y"}$)~\cite{kumar}. Further, the possibility of reconstructing the entire event allow to explore new avenues in measurements with semileptonic decays, and search for rare decays with unreconstructed particles, such as $D^0 \to \nu \bar{\nu}$, and other searches with missing energy. New flavour tagging techniques, called rest-of-event (ROE) tagging, will also enable to tag with advanced statistical methods the rest of 75\% of  \Dz mesons not tagged by the pion- and lepton-tagging techniques~\cite{casarosa}. Preliminary studies indicate that combining measurements from pion-tagged and ROE-tagged samples is equivalent to an effective increase of luminosity of about 40\%. 
Measurements of $\CP$ asymmetries on channels with charged tracks in the final state will also  be important to complement LHCb measurements, even though LHCb is expected to collect larger samples in these cases.

The LHCb experiments just ended operations at the end of 2018, after accumulating a data sample of more than 9\invfb of $pp$ collisions since its starting in 2010. This is the largest data set available for pursuing precision measurements in charm physics. All measurements presented (and also others) will be updated soon using this full sample reaching an unprecedented precision. Just to give few examples, the wrong-sign mixing analysis with $\Dz \to\Kmp\pipm$ will reach a precision of about $0.12$ on $|q/p|$ and $10^\circ$ on the mixing phase $\phi$; the measurement of $A_\Gamma$ with $\Dz\to\Kp\Km$ is expected to have an uncertainty of $0.013\%$, and that of $\Delta A_{\CP}$ with $\Dz\to h^+h^-$ decays to have an uncertainty of $0.03\%$. The LHCb experiment is now undergoing a major upgrade, called Upgrade~I, to prepare for the new LHC runs starting in 2021, where the detector will collect data at an instantaneous luminosity of $2\times10^{33}$\,cm$^{-2}$s$^{-1}$, about four time more than that of the runs just finished. All the front-end electronics and most of the subdetectors will be replaced in the next two years to allow a data-acquisition rate of 40 MHz and integrate about 50\invfb of data by 2029. Continuing on the analyses examples given above, this copious data sample will enable to have precisions of $0.03$ on $|q/p|$ and of $3^\circ$ on $\phi$, of $0.0035\%$ on $A_\Gamma$, and of $0.01\%$ on  $\Delta A_{\CP}$~\cite{LHCb-2-physics_case}.
 Consolidation and modest enhancements of the Upgrade~I detector will occur around 2025, when Belle~II is expected to accumulate is full data sample of 50\invab and end operations. It is thus highly likely that LHCb will be the only large-scale flavour-physics experiment after that time. To fully profit from the High-Luminosity LHC machine, a second major upgrade of the detector, Upgrade~II, is proposed to be installed in the long shutdown 4 of the LHC (2030), to build on the strengths of the current LHCb experiment and the Upgrade~I. It will operate at a luminosity up to $2\times10^{34}$\,cm$^{-2}$s$^{-1}$, ten times that of the Upgrade~I detector. An Expression Of Interest proposing Upgrade~II was submitted in February 2017 to the LHCC attention~\cite{LHCb-2-LoI}, and the physics case for the Upgrade~II is presented in depth in Ref.~\cite{LHCb-2-physics_case}. 

%% file: conclusion.tex
\section{Conclusion}
\label{sect:conclusion}
We presented the results on charm physics discussed in the Working Group 7 at the last CKM Workshop held at the University of Heidelberg. Thank to the constant effort of both the theoretical and experimental communities, good progress have been reported, and we could expect to have first evidence of \CP violation in the charm sector soon. 

Huge samples collected by the experiments are posing hard challenges in the analyses of data to control systematic uncertainties; anyhow, refined techniques and new methods have been presented and an unprecedented precision have been achieved in measurements of both mixing and \CP violation, further constraining the parameters space for new physics models. Among many others, we highlight new high-precision measurements  reported by LHCb in the wrong-sign mixing analysis and for the mixing-parameter $y_{\CP}$. Despite several years from the end of their operations, $B$ factories are still producing interesting new results to complement the LHCb fruitful program, {\it e.g.} the study of triple-product \CP-violating asymmetries and that of radiative charm decays, which have been suggested as sensitive probes for non-SM physics. This workshop saw also the first physics plots coming from the Belle II collaboration, showing the readiness of the experiment to enter the game. The amount of data collected by LHCb opened new avenues also in the study of charm baryons and rare $D$ decays for \CP violation measurements.  

On the theoretical side, advancements to cope with non-perturbative QCD effects making use of experimental data are on the way, as well as the understanding of subtle effects, such as those induced by kaon mixing in $D$ decays to final states with neutral kaons. All of these is pivotal to identify a genuine new physics contribution in measurements at high-precision. New avenues to search for direct \CP violation in the experiments to come have also been proposed. 

Good prospects are ahead of us, with LHCb getting ready for the Upgrade~I to start in 2021, while the Belle~II collaboration is completing the detector with the installation of the full vertex detector. Plans for a second upgrade of LHCb in 2030 to fully exploit the HL-LHC potential has been also reported.  

\section{Acknowledgements}
We would like to thank all the speakers for the excellent presentations in the WG7. We also thank the organizers of CKM2018 for giving us the opportunity to convene this fruitful working group. We really enjoyed the hospitality and the scientific ambience of the workshop. VB would like to thank SERB (India) and DST (India) for providing financial support from ITS and INSPIRE Faculty grant.